\documentclass[11pt]{article}
\usepackage{amsbsy}
\newif\iffigs\figstrue

\usepackage{epsfig,latexsym}
\usepackage{amsmath}
\usepackage{verbatim}
\usepackage{mathrsfs}
\usepackage{amssymb}

\DeclareMathAlphabet{\mathpzc}{OT1}{pzc}{m}{it}

 \csname
@addtoreset\endcsname{equation}{section}

\def\gz0{\gamma^{0}}

\def\ft#1#2\frac{{\textstyle{{\scriptstyle #1}
}{ {\scriptstyle #2}}}}

\def\beq{\begin{equation}}
\def\eeq{\end{equation}}
\def\bea{\begin{eqnarray}}
\def\eea{\end{eqnarray}}
\def\ba{\begin{array}}
\def\ea{\end{array}}
\def\bec{\begin{center}}
\def\ec{\end{center}}
\def\ba{\begin{align}}
\def\ena{\end{align}}
\def\ft{\footnote}

\def\12{\frac{1}{2}}

\begin{document}

\begin{flushright}
{\today}
\end{flushright}

\vspace{10pt}

\begin{center}


{\Large\sc $AdS$ Vacua from Dilaton Tadpoles and Form Fluxes}\\


\vspace{25pt}
{\sc J.~Mourad${}^{\; a}$  \ and \ A.~Sagnotti${}^{\; b}$}\\[15pt]

{${}^a$\sl\small APC, UMR 7164-CNRS, Universit\'e Paris Diderot -- Paris 7 \\
10 rue Alice Domon et L´eonie Duquet \\75205 Paris Cedex 13 \ FRANCE
\\ }e-mail: {\small \it
mourad@apc.univ-paris7.fr}\vspace{8pt}

{${}^b$\sl\small
Scuola Normale Superiore and INFN\\
Piazza dei Cavalieri, 7\\ 56126 Pisa \ ITALY \\
e-mail: {\small \it sagnotti@sns.it}}\vspace{10pt}

\vspace{36pt} {\sc\large Abstract}\end{center}
\noindent
We describe how unbounded three--form fluxes can lead to families of $AdS_3 \times S_7$ vacua, with constant dilaton profiles, in the $USp(32)$ model with ``brane supersymmetry breaking'' and in the $U(32)$ 0'B model, if their (projective--)disk dilaton tadpoles are taken into account. We also describe how, in the $SO(16) \times SO(16)$ heterotic model, if the torus vacuum energy $\Lambda$ is taken into account, unbounded seven--form fluxes can support similar $AdS_7 \times S_3$ vacua, while unbounded three--form fluxes, when combined with internal gauge fields, can support $AdS_3 \times S_7$ vacua, which continue to be available even if $\Lambda$ is neglected. In addition, special gauge field fluxes can support, in the $SO(16) \times SO(16)$ heterotic model, a set of $AdS_{n}\times S_{10-n}$ vacua, for all $n=2,..,8$. String loop and $\alpha'$ corrections appear under control when large form fluxes are allowed.

\setcounter{page}{1}

\pagebreak

\newpage

\baselineskip=20pt
\section{\sc  Introduction and Setup}\label{sec:intro}

Supersymmetry breaking appears typically accompanied, in String Theory \cite{strings}, by the emergence of runaway potentials. The vacuum is deeply affected by their presence and current string tools, which are very powerful when combined with supergravity \cite{SUGRA} in the supersymmetric case, become ineffective in front of the resulting redefinitions. These reflect an untamed problem that started to surface long ago \cite{fs}.

As in other systems, internal fluxes can make a difference in this context, and interesting progress was recently made in~\cite{closed_fluxes}, while~\cite{ang_card_irges} contains notable earlier results. What we shall present here bears some similarities to these works, although it originates from a different line of thought, as we are about to describe. In this letter, relying on the low--energy effective field theory, we explore vacuum solutions in ten--dimensional non--tachyonic models, taking into account the low--lying potentials that are induced when supersymmetry is broken or absent altogether. The cases at stake include two types of orientifold models \cite{orientifolds} and the $SO(16) \times SO(16)$ heterotic model. In the first orientifold model, an exponential potential results from the (projective--)disk tadpoles that accompany ``brane supersymmetry breaking'' \cite{bsb}. This string--scale mechanism reflects the presence, in vacua that host a non--linear realization of supersymmetry \cite{dmnlin}, of the residual tension from non--BPS combinations of BPS branes and orientifolds. The simplest example of this type is Sugimoto's model in \cite{bsb}. Its low--lying closed sector includes the graviton, a dilaton, a RR two--form potential, a Majorana-Weyl gravitino and a Majorana--Weyl spinor of opposite chiralities, as in the type-I superstring. However, in the low--lying open sector there are adjoint vectors of a $USp(32)$ gauge group and fermions in the reducible antisymmetric representation, whose singlet plays the role of a goldstino. There is also another non--tachyonic orientifold, which is not supersymmetric to begin with but whose features are rather similar. It is the 0'B model of \cite{0bprime}, whose closed sector is purely bosonic and contains, in addition to the bosonic modes of the type-I superstring, an axion and a self--dual four form potential, and whose open sector hosts the vectors of a $U(32)$ gauge group and Fermi modes in the antisymmetric representation and its conjugate. Both settings afford a rich family of lower--dimensional counterparts \cite{bsb,0bprimecarlo}, but here we shall content ourselves with the simplest 10D examples. Finally, in the third non--tachyonic 10D non--supersymmetric model, of the heterotic type \cite{so16so16}, the bosonic spectrum describes again gravity, a dilaton and a two--form potential that comes from the NS-NS sector, together with $SO(16) \times SO(16)$ gauge fields. These theories are anomaly free, but the 0'B model rests on a more complicated, non--factorized Green-Schwarz mechanism, as a wide class of lower--dimensional examples \cite{gs,orientifolds,0bprime}.

A 9D vacuum of Sugimoto's model was exhibited long ago by Dudas and one of us \cite{dmvac}, but it contains singularities and regions of strong coupling. In this note we shall see that form fluxes can lead to some smooth symmetric vacua for 10D orientifolds even in the presence of dilaton tadpoles. Moreover, when large form fluxes are allowed the resulting $\alpha'$ and string loop corrections appear small. There are $AdS_3 \times S_7$ solutions of this type for both the $USp(32)$ model of \cite{bsb} and the $U(32)$ 0'B model of \cite{0bprime}, and moreover there are similar $AdS_3 \times S_7$ and $AdS_7 \times S_3$ solutions for the heterotic $SO(16) \times SO(16)$ model.

The low--energy dynamics of the systems of interest is described, in the Einstein frame, by effective Lagrangians of the type
\bea
{\cal S} &=& \frac{1}{2\,k_{10}^2}\, \int d^{10}x\sqrt{-g}\left\{-\ R\ - \frac{1}{2}\ (\partial\phi)^2\ - \ \frac{1}{2\,(p+2)!}\ e^{-2\,\beta_E^{(p)}\,\phi}\, {\cal H}_{p+2}^2 \right. \nonumber \\
&-& \left. \frac{1}{4} \ e^{-2\,\alpha_E\,\phi}\, {\rm tr}\, {\cal F}^2 \ - \ T \, e^{\,\gamma_E\,\phi} \,
\right\} \ , \label{lagrangian}
\eea
where $2\,k_{10}^2=(2\pi)^7$ and we have set $\alpha'=1$. These originate from the string--frame actions
 \bea
 {\cal S}&=&\frac{1}{2\,k_{10}^2}\ \int d^{10}x\,\sqrt{-\,G}\left\{ e^{-2\phi}\left[\, - \,R\, + \,4(\partial\phi)^2 \right]\, - \,\frac{1}{2(p+2)!}\ e^{-2\,\beta_S \,\phi}\ {\cal H}_{p+2}^2\right.
 \nonumber \\ &-& \left. \frac{1}{4}\ e^{-\,2\,\alpha_S\,\phi}\, {\rm tr}\, {\cal F}^2 \,-\,T \ e^{\,\gamma_S\,\phi} \, \right\} \ , \label{eqs1}
 \eea
and ${\cal H}_{p+2}$ is the field strength of a $(p+1)$--form potential ${\cal B}_{p+1}$~\footnote{The definition of the ${\cal H}_{p+2}$'s involves, in general, Chern--Simons terms related to the Green--Schwarz cancellation mechanism, or its generalization, at work in these systems.} that presents itself, in various incarnations, in both heterotic and orientifold 10D models. Moreover $\gamma_S=-1$ for the orientifold models, where the contribution arises from \emph{(projective) disk} amplitudes, while $\gamma_S=0$ for the $SO(16) \times SO(16)$ heterotic model, where the modification arises from the \emph{torus} amplitude. Hence, $\gamma_E=\frac{3}{2}$ for orientifold models and $\gamma_E=\frac{5}{2}$ for the $SO(16) \times SO(16)$ heterotic model. In this last case a symbol $\Lambda$ would be more appropriate, but in eqs.~\eqref{lagrangian} and \eqref{eqs1} we have used $T$ in all cases for brevity.
The parameters $\beta_E^{(p)}$ that enter eqs.~\eqref{lagrangian} and \eqref{eqs1} are related by Weyl rescalings to their string--frame counterparts, and $\beta_S=1$ for the heterotic NS-NS two--form potential of ordinary 10D supergravity \cite{10dSUGRA} while $\beta_S=0$ for the $R-R$ two--form potential present in orientifold models \cite{wittenM}.
Moreover, $\alpha_E= -\frac{1}{4} \ (\alpha_S= \frac{1}{2})$ for the orientifold models and $\alpha_E= \frac{1}{4} \ (\alpha_S= 1)$ for the $SO(16)\times SO(16)$ model.

Our aim here is to exhibit for the three 10D models solutions of the type
\beq
ds^{\,2}\ = \ e^{2A(r)}\, g(\mathtt{L_k})\ + \ dr^2\ + \ e^{2C(r)}\, g(\mathtt{E_{k'}})\ , \label{metric}
\eeq
where $\mathtt{L_k}$ is a maximally symmetric $(p+1)$--dimensional Lorentzian manifold of curvature $k$, while  $\mathtt{E_{k'}}$ is a maximally symmetric $(8-p)$--dimensional Euclidean manifold of curvature $k'$. These space times possess a manifest $ISO(1,p) \times SO(9-p)$ symmetry when $k=0$ and $k'=1$, which will be the case of main interest for us \footnote{If $k'=-1$, the manifest internal isometry would be $SO(1,8-p)$. Moreover, the space--time portion would have a manifest $SO(1,p+1)$ isometry if $k=1$ and a manifest $SO(2,p)$ isometry if $k=-1$.}, and rest, in general, on two scalar functions, $A(r)$ and $C(r)$. This is a convenient setting to explore brane configurations, if one allows for non--trivial profiles of the dilaton and the available form-fields. Non--abelian gauge fields valued in suitable subalgebras of the gauge algebra can also be compatible with the manifest isometries. These properties are granted by any radial profile $\phi(r)$ for the scalar, by arbitrary scale factors $A(r)$ and $C(r)$, and by form-fields proportional to the $(p+1)$--dimensional volume element, up to a second scalar function $b(r)$. Here, however, we shall restrict our attention to highly symmetric configurations where both $\phi$ and $C$ are constant.
For gauge fields, non--trivial symmetric profiles exist in this case for $k'=1$. The resulting internal spaces are spheres, whose $SO(9-p)$ isometries can be identified with corresponding orthogonal subgroups of the available gauge groups, $USp(32)$, $U(32)$ or $SO(16) \times SO(16)$, up to a third scalar function $a(r)$. In detail, given the embedding coordinates $\tilde{y}^i$ on the internal spheres, such that $\tilde{y}^T \, \tilde{y}=1$, the $so(9-p)$--valued gauge field configurations read
\beq
{\cal A} \ = \ i\, a(r) \left( \tilde{y}\, d\tilde{y}^T \ - \ d\tilde{y}\, \tilde{y}^T \right) \ . \label{gaugeprofile}
\eeq
As we shall see shortly, the solutions resulting from this set of profiles for the different available fields, when they exist, describe vacua of the $AdS \times S$ type.

A common signature of the 10D $USp(32)$ and $U(32)$ orientifold models is the special value $ \gamma_E \ = \ \frac{3}{2}$,
which reflects the (projective--)disk origin of the dilaton potential. In both cases one can turn on ${\cal H}_{p+2}$ form fluxes with $p=1$ or $p=5$, corresponding to two--form potentials or to dual six--form ones, and the preceding discussion implies that
\beq
\alpha_E \ = \ - \ \frac{1}{4}\ , \qquad \beta_E^{(1)} \ = \ - \ \frac{1}{2}  \ , \qquad \beta_E^{(5)} \ = \ \frac{1}{2} \ . \label{abc_orientifold_USp}
\eeq
Moreover, the $U(32)$ model also allows profiles of a four--form potential with self--dual field strength and of an eight--form dual potential, for which
\beq
\beta_E^{(3)} \ = \ 0 \ , \qquad  \beta_E^{(7)} \ = \ 1 \ . \label{abc_orientifold_U}
\eeq
Finally, as we have anticipated, in the heterotic $SO(16) \times SO(16)$ model $\gamma_E \ = \ \frac{5}{2}$,
since the dilaton potential originates from the torus level, and one can turn on form fluxes with $p=1$ or $p=5$, with
\beq
\alpha_E \ = \ \frac{1}{4}\ , \qquad \beta_E^{(1)} \ = \ \frac{1}{2} \ , \qquad \beta_E^{(5)} \ = \ - \ \frac{1}{2} \ . \label{abc_heterotic}
\eeq
\section{\sc $AdS \times S$ Solutions of 10D Non--Tachyonic Models}

The class of solutions that we would like to illustrate rests on constant values for $C$ and $\phi$, on special constant values for the gauge field functions $a$ in eq.~\eqref{gaugeprofile} that solve the corresponding non--linear field equations, and on the explicit non--constant solutions
\beq
{\cal H}_{p+2} \ = \ h\, e^{\,(p+1)\,A(r) \,+\, 2\, \beta_E^{(p)}\,\phi \, -\, (8-p) \,C} \ \epsilon(p+1) \, dr
\eeq
for the relevant form field strengths, where $h$ is a constant and $\epsilon(p+1)$ denotes the $(p+1)$--dimensional volume form. With these types of profiles the field equations arising from \eqref{lagrangian} reduce to
\bea
T\, e^{\,\gamma_E\,\phi} &=& \frac{\xi\,\alpha_E}{\gamma_E} \ \left(8\,-\,p\right)\left(7\,-\,p\right)\, e^{\,-\,4\,C\,-\,2\,\alpha_E\,\phi} \ - \ \frac{\beta_E^{(p)}\,h^2}{\gamma_E} \ e^{\,-\,2\,(8\,-\,p)\,C\,+\,2\,\beta_E^{(p)}\,\phi} \, , \label{eq73} \\
16\,k'\, e^{\,-\,2\,C} &=& \xi \left[ 8+p \, + \, \frac{2\,\alpha_E}{\gamma_E} \, \left(8-p \right) \right]\, e^{\,-\,4\,C\,-\,2\,\alpha_E\,\phi} \nonumber \\ &+&  \frac{h^2 \left( p\ +\ 1 \, - \, \frac{2\,\beta_E^{(p)}}{\gamma_E} \right)e^{\,-\,2\,(8\,-\,p)\,C\,+\,2\,\beta_E^{(p)}\,\phi}}{\left(7\,-\,p\right)} \ , \label{eq83} \\
(A')^2 &=& k\,e^{\,-\,2\,A} \,+\, \xi \ \frac{(8\,-\,p)(7\,-\,p)}{16(p\,+\,1)}\  \left(1\,-\, \frac{2\alpha_E}{\gamma_E} \right)e^{\,-\,4\,C\,-\,2\,\alpha_E\,\phi} \nonumber \\
&+& \frac{h^2}{16(p\,+\,1)} \ \left( 7\,-\,p \,+ \, \frac{2\,\beta_E^{(p)}}{\gamma_E} \right)\ e^{\,-\,2\,(8\,-\,p)\,C\,+\,2\,\beta_E^{(p)}\,\phi} \ . \label{eq93}
\eea
Here $\xi=0,1$ define two distinct choices for the internal gauge field strength,
\beq
{\cal F} \ = \ i\, \xi \ d\tilde{y}\, d\tilde{y}^T \ , \label{gaugefieldstrength}
\eeq
which correspond to $a=\frac{\xi}{2}$, both of which satisfy the corresponding field equations, with ${\cal F}$ vanishing in the first case. In three dimensions the choice $\xi=1$ would identify the Wu--Yang solution \cite{wuyang}.

Notice that eq.~\eqref{eq73} implies strong constraints, due to the positivity of $T$ (or $\Lambda$, in the $SO(16) \times SO(16)$ model). Unbounded values of $h$ are manifestly possible, in the absence of internal gauge fields, when $\beta_E^{(p)}<0$, and thus for three--form fluxes in the orientifold models or seven--form fluxes in the heterotic model. In addition we shall see that, surprisingly, unbounded three--form fluxes are also allowed in the $SO(16) \times SO(16)$ model, in the presence of non--trivial internal gauge fields.

As is well known, the field strength ${\cal H}_{p+2}$ involves, in general, Chern--Simons couplings. These, in their turn, bring about a subtlety related to the Bianchi identities, which are modified into
\beq
d\,{\cal H}_{p+2} \ \sim \ X_{p+2}\left( {\cal F}, R \right) \ ,
\eeq
where the $X$'s are invariant polynomials involving space--time and gauge--field curvatures. However, in our highly symmetric backgrounds these Chern--Simons terms vanish, and the $X$'s with them, so that there are no further conditions coming from this end. For the internal gauge fields this is manifest from eq.~\eqref{gaugefieldstrength}, since $d\tilde{y}^T\,d\tilde{y}=0$, and a similar link holds between the vielbein one--form and the Riemann curvature.

With constant $\phi$ and $C$ profiles, the space--time manifold acquires additional isometries, and its $AdS$ nature is evident from eqs.~\eqref{metric} and \eqref{eq93} for $k=0$, since in this case one recovers a standard presentation of the symmetric space in Poincar\'e coordinates. Actually, when combined with the radial direction, $L_k$ is always describing an $AdS$ space, independently of the value of $k$: this only affects the slicing, which is Minkowski for $k=0$, $dS$ for $k=1$ and $AdS$ for $k=-1$.

We can now analyze in detail the solutions of eqs.~\eqref{eq73}, \eqref{eq83} and \eqref{eq93}.

\subsection{\sc  $AdS_3 \times S_7$ Solutions in 10D Orientifolds}\label{sec:BSB_AdS}

Taking into account the corresponding vacuum configurations, one can see that for Sugimoto's model $T  =  \frac{16}{\pi^2} $,
while the 0'B model the total tension is half of this value. For their $RR$ two--form potentials $\beta_E$ is negative, which allows unbounded values for $h$. Since for these models $\alpha_E<0$, as we have seen in eq.~\eqref{abc_orientifold_USp}, the very consistency of eq.~\eqref{eq73} requires a non--vanishing $h$. As a result, the vacua that we are about to describe are sustained by three--form fluxes that, strictly speaking, are quantized but can be unbounded.
In addition, the \emph{r.h.s.} of eq.~\eqref{eq73} is manifestly non-negative, and thus $k'=1$, so that the internal spaces of these solutions are seven--dimensional spheres $S_7$. One has also the option of turning on internal gauge fields, provided $h$ is not vanishing. The solutions that we have identified are thus $AdS_3 \times S_7$ vacua.

The key features of this class of solutions become quite transparent if $\xi=0$, but in general eqs.~\eqref{eq73} and \eqref{eq83} imply that, for the two non--tachyonic orientifold models,
\bea
e^{\,2\,C} &=&  \frac{2\,\xi\,e^{\,\frac{\phi}{2}}}{1 \ \pm \ \sqrt{1 \ - \ \frac{\xi\,T}{3}\, e^{\,2\,\phi}}} \nonumber \\
\frac{h^2}{32} &=& \frac{ \xi^7 \, e^{\,4\,\phi}}{\left(1 \ \pm \ \sqrt{1 \ - \ \frac{\xi\,T}{3}\, e^{\,2\,\phi}} \right)^7} \left[ \frac{42}{\xi}\, \left(1 \ \pm \ \sqrt{1 \ - \ \frac{\xi\,T}{3}\, e^{\,2\,\phi}} \right) \ +\ {5} \ T\, e^{\,2\,\phi} \right] \ . \label{Rgs_orientifold}
\eea
For the reader's convenience, we have kept all powers of $\xi$, despite the fact that $\xi=0,1$, in order that the smooth limiting behavior of these and other solutions as $\xi \to 0$ be manifest.
There is a branch of solutions, corresponding to the ``minus'' sign above, which connects smoothly to the $\xi=0$ case, where large $AdS_3$ and $S_7$ radii accompany small couplings. In this case the preceding relations imply
the $\xi$--independent limiting large--$h$ behavior
\beq
g_s \ \equiv \ e^{\,\phi} \ \sim \ \frac{12}{\left(2\,h\,T^3\right)^\frac{1}{4}} \ , \qquad R^4  \, g_s^{3} \ \sim \ \frac{144}{T^2} \ , \qquad \left(A'\right)^2 \ \sim \ k \, e^{\,-\,2\,A} \ + \ \frac{{6}}{R^2} \ . \label{eq103}
\eeq
Notice the crucial role played by the tension $T$ in granting the existence of these solutions.
For large values of $h$ these expressions receive small corrections also for $\xi=1$, and thus within an interesting corner of parameter space the complete equations of String Theory appear reliably captured by our approximations. Large values of $h$ imply large values for $R$, which sets the scales for the $AdS_3$ and $S_7$ factors, and also small values for the string coupling. These, in their turn, are expected to translate into small $\alpha'$ and string loop corrections. Finally, for $\xi=0$ the $USp(32)$ gauge group is unbroken, while for $\xi=1$ it is broken to a $USp(24)$ subgroup.

The second branch of solutions, corresponding to the ``plus'' sign in eq.~\eqref{Rgs_orientifold}, is only available for $\xi=1$, and thus in the presence of internal gauge fields. It associates large radii to strong couplings (whose upper bound is determined by $T$) and, surprisingly, has a smooth limit for vanishing tension $T$, when it is also a solution for the $SO(32)$ type-I superstring \cite{gs}. Although it lies outside the perturbative reach, we find this option intriguing, especially in view of a similar weak--coupling heterotic counterpart that we are about to describe.

These considerations also apply to the 0'B orientifold, where if $\xi=0$ the original $U(32)$ gauge group is unbroken, while if $\xi=1$ it is broken to a $U(24)$ subgroup. In this case there would be also, in principle, the two other options of eq.~\eqref{abc_orientifold_U}, which correspond to $p=3,7$. However, they do not lead to consistent solutions of this type, since the corresponding values for $\beta_E$ are not positive, while the corresponding $\alpha_E$ of eq.~\eqref{abc_orientifold_USp} are negative.

\subsection{\sc  $AdS_3 \times S_7$ and $AdS_7 \times S_3$ Solutions in the 10D Heterotic $SO(16) \times SO(16)$ }\label{sec:nosusy_AdS}

In the heterotic $SO(16) \times SO(16)$ model the sign of $\alpha_E$ is positive, as we have seen in eq.~\eqref{abc_heterotic}. As a result, internal non--abelian gauge fields can sustain by themselves the class of vacua under scrutiny. Let us also recall that for this model $\Lambda \simeq \frac{4\pi^2}{25}$ \cite{so16so16}.

If only internal gauge fields are retained, eq.~\eqref{eq83} reveals that $k'=1$, so that the internal spaces are again, consistently, spheres. In the absence of form fluxes there is thus a whole family of $AdS_{p+2} \times S_{8-p}$ solutions, for $p=1,...,6$, where the original gauge group is broken accordingly. The problem with these solutions is that the value of $a$ that solves the gauge field equations is fixed, so that one looses the large deformation parameter available in the preceding examples. As a result, these solutions are expected to suffer from sizable $\alpha'$ and string loop corrections.

For $p=1$, however, one can also turn on a three--form flux, but $\beta_E$ is now positive and naively this would seem to imply bounded values for $h$. However, the actual range for $h$ is the result of competing effects, and definite assessments can be made only after combining eqs.~\eqref{eq73} and \eqref{eq83}. In this fashion one can show that
\beq
e^{\,2\,C} \ = \ \frac{2\, \xi\, e^{\,-\,\frac{\phi}{2}}}{1 \ + \ \sqrt{1 \ + \ \frac{\xi\,\Lambda}{3}\, e^{\,2\,\phi}}} \ , \label{eq9Ch}
\eeq
since only one branch is compatible with the positivity of the exponential of $C$.  Notice that these solutions are supported by the internal gauge fields and disappear for $\xi=0$, and letting $\xi=1$ the corresponding link between $h$ and $\phi$ reads
\beq
\frac{h^2}{32} \ = \  \frac{e^{\,-\,4\,\phi}}{\left[ 1 \ + \ \sqrt{1 \ + \ \frac{\Lambda}{3}\, e^{\,2\,\phi}}\right]^7 }\ \left[42 \left( 1 \ + \ \sqrt{1 \ + \ \frac{\Lambda}{3}\, e^{\,2\,\phi}}\right) \ - \ 13\, \Lambda\, e^{\,2\,\phi} \right] \ . \label{eq9ph}
\eeq

The preceding relations encode the limiting large--$h$ behavior
\beq
g_s \ \equiv \ e^{\,\phi} \ \sim \ \left(\frac{21}{h^2}\right)^\frac{1}{4} \ , \qquad g_s\, R^4 \ \sim 1 \ , \qquad \left(A'\right)^2 \ \sim \ k \, e^{\,-\,2\,A} \ + \  \frac{21}{4\,R^2} \ , \label{limit_het_37}
\eeq
so that even this class of solutions affords a region where $\alpha'$ and string loop corrections appear under control. Notice that these weak--coupling solutions of eqs.~\eqref{eq9Ch} and \eqref{eq9ph} have a smooth limit for vanishing $\Lambda$, which is not needed to grant their existence. The marginal role of $\Lambda$ in this case is clearly reflected in the limiting behavior of eqs.~\eqref{limit_het_37}, and indeed this class of solutions would be available also in the supersymmetric 10D heterotic models of \cite{ghmr}. This contrasts with what we found for first branch of $AdS_3 \times S_7$ solutions in the orientifold models, where a (projective--)~disk tadpole was essential, but it resonates with what we found for the second branch, whose strong--coupling solutions also continue to exist in the limit of vanishing $T$. On the other hand, a strong--weak coupling link is expected to hold when both $T$ and $\Lambda$ vanish, between solutions of the $SO(32)$ type-I superstring and of the $Spin(32)/Z_2$ heterotic \cite{ghmr} model, which are weak--strong coupling partners in the general picture of \cite{wittenM}. An extension of the link to the case of non--vanishing $T$ and $\Lambda$, which the solutions somehow suggest, could provide interesting clues on non--supersymmetric string dualities~\footnote{The link that is surfacing here is related to heterotic--type I duality along the lines of \cite{wittenM}. Other dualities relating heterotic strings with broken and unbroken supersymmetry, suggested by Scherk--Schwarz deformations~\cite{scherk_schwarz}, were first explored long ago in~\cite{blum_dienes}.}.

This is not all for this heterotic $SO(16) \times SO(16)$ model, since there is another potentially interesting flux, corresponding to $p=5$, and thus to the six--form potential present in the dual formulation of \cite{chamseddine}. The duality connecting this case to the more familiar one with $p=1$ entails a reversal of the sign of $\beta_E$, as we have seen, so that the resulting vacua can again involve unbounded tensor profiles, possibly accompanied by internal gauge fields as in eqs.~\eqref{gaugeprofile} and \eqref{gaugefieldstrength}. The gauge group is again unbroken in their absence, and otherwise it is broken to $SO(16) \times SO(12)$.

Let us take a closer look at this last class of solutions. Combining eqs.~\eqref{eq73} and \eqref{eq83}, one can now derive the two relations
\bea
e^{\,2\,C} &=&  \frac{\xi}{2}\ \frac{e^{\,-\,\frac{\phi}{2}}}{1 \ - \ \sqrt{1 \ - \ {\xi\,\Lambda}\, e^{\,2\,\phi}}} \ ,\label{Rgs_p5_heterotic} \\
\frac{h^2}{3} &=& \ {\xi^3}\ \frac{ \left[ \frac{17\,\Lambda}{24}\,e^{\,2\,\phi}\ - \ \frac{1}{\xi}\  \left(1 \ - \ \sqrt{1 \ - \ {\xi\,\Lambda}\, e^{\,2\,\phi}} \right) \right]}{\left(1 \ - \ \sqrt{1 \ - \ {\xi\,\Lambda}\, e^{\,2\,\phi}} \right)^3} \ , \nonumber
\eea
and there is again a single consistent branch, which now connects smoothly to the $\xi=0$ case. These expressions clearly show that this last class of solutions rests on $\Lambda$, and ceases to exist in its absence. Finally, in the large--$h$ limit
\beq
g_s \ \equiv \ e^{\,\phi} \ \sim \ \left(\frac{5}{h^2\,\Lambda^2}\right)^\frac{1}{4} \ , \qquad g_s^5 \, R^4 \ \sim \ \frac{1}{\Lambda^2}  \ , \qquad \left(A'\right)^2 \ \sim \ k \, e^{\,-\,2\,A} \ + \ \frac{1}{4\,R^2}  \ . \label{eq107}
\eeq

Once more, large values for $h$ result in large $AdS_7$ and $S_3$ radii, and also in small values for the string coupling $g_s$. These results are expected to translate into small $\alpha'$ and string loop corrections.

%
\section{\sc  Conclusions}\label{sec:conclusions}

We have described how runaway exponential potentials arising from (projective--)disk dilaton tadpoles can combine with form fluxes to yield a family of $AdS_3 \times S_7$ vacua for Sugimoto's orientifold model of \cite{bsb} and for the non--tachyonic 0'B orientifold of \cite{0bprime}, with unbroken gauge groups. Alternatively, if internal non--abelian profiles are also turned on the gauge groups break, in the two cases, to $USp(24)$ and $U(24)$. We have shown that there are actually two branches of solutions. The first branch is a more conventional weak--coupling one, which can be supported by a three--form flux alone and whose existence rests on the presence of the tadpole tension $T$. The second branch is a strong--coupling one, which rests on the presence of internal gauge fields and continues to exist in the limit of vanishing tension, when it applies to the $SO(32)$ type-I superstring.
In the $SO(16) \times SO(16)$ heterotic model \cite{so16so16} internal gauge fields alone can sustain $AdS_n \times S_{10-n}$ vacua for $n=2,..,8$, but perturbative large--field limits exist only when form fields can be included, \emph{i.e.} for $n=3,7$. In the first case the vacuum is sustained by the essential contribution of internal gauge fields, combined with an ${\cal H}_3$ flux, and continues to exist in the limit of vanishing $\Lambda$, when the solution would also apply to supersymmetric heterotic strings, consistently with the string duality link between the $Spin(32)/Z_2$ heterotic string and the $SO(32)$ type-I superstring. If a link transcends the case of vanishing $T$ and $\Lambda$, as the solutions would seem to suggest, it encodes potentially interesting information of string dualities beyond the supersymmetric case, on the par with cases explored long ago in \cite{blum_dienes}. We have also found a second class of weak--coupling solutions of the $SO(16) \times SO(16)$ heterotic model, supported by ${\cal H}_7$ fluxes, which exists only in the presence of a non--vanishing $\Lambda$.

Although our results rest on the low--energy field equations captured by (super)gravity, in the $AdS_3 \times S_7$ orientifold solutions and in both the $AdS_3 \times S_7$ and $AdS_7 \times S_3$ heterotic solutions one can turn on large form fluxes, which result in large manifold sizes and small string couplings, so that both $\alpha'$ and string loop corrections are expected to be small in these regions of parameter space.

We were led to the present considerations by our interest in the general problem of back--reactions to broken supersymmetry in String Theory, and in particular to the string--scale phenomenon of ``brane supersymmetry breaking'' \cite{bsb}. The simple vacua that we have found have the virtue of lacking spatial regions of strong coupling, which were present in the nine--dimensional one found in \cite{dmvac}, and a variety of options of this type can be available in lower--dimensional cases.

The 0'B model was explored by Armoni and others \cite{armoni}, over the years, with an eye to large--$N$ limits of $QCD$, in view of its formal proximity to a supersymmetric spectrum. The vacua that we have exhibited might prove useful for this program, or in other directions related to the AdS/CFT correspondence \cite{adscft}. In addition, one could consider compactifications to four dimensions starting from a Randall--Sundrum extension \cite{rs} of the $AdS_7 \times S_3$ heterotic solution.

It would be interesting to explore charged brane solutions of the orientifold models that we have considered, and also their uncharged counterparts that can be built following Sen \cite{sen}. It would be very interesting, in particular, to investigate how the CFT spectra of \cite{dms}, which were determined by world--sheet techniques, and thus ignoring tadpole potentials, are affected by the deformed backgrounds. These problems and other stability issues are currently under scrutiny.
\vskip 24pt
\section*{Acknowledgments}

\vskip 12pt

We are grateful to E.~Dudas and G.~Pradisi for their collaboration on related projects, and to E.~Dudas for stimulating discussions. AS is grateful to Scuola Normale and INFN (IS Stefi) for partial support, and to
APC---U.~Paris VII for partial support and the kind hospitality. Finally, we would like to thank the referee for calling to our attention some references that we had inadvertently ignored.

\end{document}